\documentclass[twocolumn,english,nofootinbib]{revtex4}
\usepackage[T1]{fontenc}
\usepackage[latin9]{inputenc}
\usepackage{array}
\usepackage{float}	
\usepackage{amsmath}
\usepackage[pdftex]{graphicx}
\usepackage{amssymb}
\usepackage{color}
\usepackage{comment}

\makeatletter

\providecommand{\tabularnewline}{\\}

\@ifundefined{textcolor}{}
{%
 \definecolor{BLACK}{gray}{0}
 \definecolor{WHITE}{gray}{1}
 \definecolor{RED}{rgb}{1,0,0}
 \definecolor{GREEN}{rgb}{0,1,0}
 \definecolor{BLUE}{rgb}{0,0,1}
 \definecolor{CYAN}{cmyk}{1,0,0,0}
 \definecolor{MAGENTA}{cmyk}{0,1,0,0}
 \definecolor{YELLOW}{cmyk}{0,0,1,0}
 }

\makeatother

\usepackage[normalem]{ulem}  

\newcommand{\dsp}{\displaystyle}
\newcommand{\MeV}{{\rm MeV}}

\newcommand{\eqn}[1]{(\ref{#1})}
\usepackage{babel}

\begin{document}

\title{Large amplitude behavior of the bulk viscosity of dense matter}

\author{Mark G. Alford, Simin Mahmoodifar and Kai Schwenzer}

\affiliation{Department of Physics, Washington University, St. Louis, Missouri,
63130, USA}

\begin{abstract}
We study the bulk viscosity of dense matter, taking into account
non-linear effects which arise in the large amplitude ``supra-thermal''
region where the deviation $\mu_\Delta$ of the chemical potentials 
from chemical equilibrium fulfills $\mu_\Delta\gtrsim T$. 
This regime is relevant to
unstable modes such as r-modes, which grow in amplitude until saturated
by non-linear effects.
We study the damping due to
direct and modified Urca processes in hadronic matter, and
due to nonleptonic weak interactions in strange quark matter.
We give general results valid for an
arbitrary equation of state of dense matter and find that the viscosity
can be strongly enhanced by supra-thermal effects. 
Our study confirms previous
results on quark matter and shows that the non-linear enhancement
is even stronger in the case of hadronic matter. Our results can be applied
to calculations of the r-mode-induced spin-down of fast-rotating neutron stars,
where the spin-down time will depend on the saturation amplitude of the r-mode.
\end{abstract}
\maketitle

\section{Introduction}

Compact stars are the only known objects that contain equilibrated
matter that is compressed beyond nuclear density, making them a valuable
laboratory for the study of the structure of matter under extreme
conditions. In addition to hadronic matter they may
also contain new forms of matter that involve deconfined quarks \cite{Itoh:1970uw,Bodmer:1971we,Witten:1984rs,Alford:2007xm}.
In contrast to the static properties of compact stars which only depend
on the equation of state  of matter \cite{Lattimer:2000nx}, dynamic
properties also depend on the low energy degrees of freedom and thereby
might be able to discriminate more efficiently between different forms
of strongly interacting matter. One of the dynamic properties of
dense matter is viscosity, which determines the damping of mechanical
perturbations, and a particularly important application is to the damping
of r-mode oscillations of compact stars
\cite{Papaloizou:1978,Andersson:1997xt,Friedman:1997uh,Lindblom:1999yk,Madsen:1999ci},
which, at sufficiently low viscosity and high rotation rate,
are unstable and 
can cause rapid spin-down of the star \cite{Owen:1998xg} via gravitational
radiation.
Since the r-mode is unstable, its exponential growth must eventually
be stopped by some non-linear mechanism. Finding the relevant mechanism
is important because it determines the amplitude at which the r-mode
saturates, and hence the rate at which it spins down the star.
Previously suggested
mechanisms include mode coupling and the transformation of the
r-mode energy into differential rotation \cite{Lindblom:2000az,Gressman:2002zy,Lin:2004wx,Bondarescu:2007jw}, and friction between different layers of 
the star such as ``surface rubbing'' at the crust \cite{Bildsten:2000}. 
Because of the complexity of the problem,
these mechanisms have to rely on approximations that are not always
well controlled.

In this paper we consider an alternative mechanism which does not involve
additional physics, but is already present
in a quasi-static hydrodynamic description.
At low amplitudes the bulk viscosity is amplitude-independent, but
since the r-mode is unstable its amplitude grows, and unless 
stopped by other mechanisms will quickly enter the
``supra-thermal'' regime where the bulk viscosity grows with amplitude,
and may become large enough to stop the growth of the mode
\cite{Madsen:1992sx,Reisenegger:2003pd}.
The supra-thermal regime is characterized by $\mu_\Delta\gtrsim T$,
where $T$ is the temperature
and $\mu_\Delta$  is the amplitude of the oscillations in the chemical potential
of the quantity whose re-equilibration causes the viscous damping.
We will study the microscopic part of the
problem via a comprehensive analysis of the bulk viscosity of dense
matter. We leave the astrophysical aspects for future work. Since
the precise phase structure and the equation of state of matter at
high density is still unknown, we keep the dependence on those
parameters as explicit as possible, and as well as numeric results
we provide analytic approximations which prove
to be surprisingly accurate. This allows us to obtain general results for
the bulk viscosity valid for many different phases of matter, and enables us to
estimate the involved uncertainties. We study in detail the cases
of equilibrated $npe$-matter and strange quark matter,
and consider both modified and direct Urca processes in
the hadronic case. Yet, our general expressions can be applied to
other equations of state and entirely different forms of strongly
interacting matter.

\section{Bulk viscosity of dense matter\label{sec:general-viscosity}}

The bulk viscosity of a given form of matter is a measure of
the energy dissipated when it is subjected to
an oscillating cycle of compression and rarefaction.
Bulk viscosity is known to be the dominant source for the
damping of r-mode oscillations at high temperatures and low
amplitudes. Consequently, bulk viscosity has been computed for many
forms of dense matter \cite{Sawyer:1989dp,Haensel:2000vz,Haensel:2001mw-ADS,
Chatterjee:2007qs,Gusakov:2007px,Chatterjee:2007ka,Lindblom:2001hd,Jones:2001ya,Haensel:2001em-ADS,Chatterjee:2007iw-ADS,Madsen:1992sx,Manuel:2004iv,Alford:2006gy,Sa'd:2006qv,Alford:2007pj,Alford:2007rw,Manuel:2007pz,Sa'd:2007ud,Dong:2007ax,Sa'd:2009vx}.
Nearly all these studies restricted themselves to the sub-thermal
regime $\mu_\Delta \ll 2\pi T$. However, as noted above, the astrophysically
interesting scenario is one where the r-mode is unstable, and so
the supra-thermal bulk viscosity may well become relevant.
The influence
of the supra-thermal regime has been studied numerically in 
\cite{Madsen:1992sx}
for the case of strange quark matter. This analysis showed
that for large amplitude oscillations the viscosity can increase by
orders of magnitude. Yet, because of their qualitatively different low
energy degrees of freedom and weak-interaction equilibration
channels, other forms of matter could show different behavior.

In the following we will derive the non-linear equations that determine
the bulk viscosity due to weak interactions that interconvert the
fermionic species that are present\footnote{We do
not study bulk viscosity arising from
the interconversion of bosons, such as
the light mesons that occur in color-flavor-locked phases.}.
We will then solve it for arbitrary
amplitudes. We focus on weak interactions because their equilibration
rate is comparable to typical compact star oscillation frequencies:
strong interactions make a negligible contribution at these frequencies
because their equilibration rate is much too fast.


The bulk viscosity of a given form of matter is defined by the
response of the system to an oscillating compression and rarefaction. 
This corresponds to an oscillation in the densities of all
exactly conserved quantities. We will assume that there is at least
one such quantity whose density we call $n_*$.
In compact stars it is typically the baryon number density.
We will study the energy dissipated as a result of
a small harmonic oscillation $\delta n_{*}$ 
around its equilibrium value $\bar n_{*}$
\begin{equation}
n_{*}(\vec r,t) = \bar{n}_{*}(\vec r)+\delta n_{*}(\vec r,t)
 =\bar{n}_{*}(\vec r)+\Delta\! n_{*}(\vec{r})\sin\left(\frac{2\pi t}{\tau}\right)\,,
\end{equation}
where the amplitude of the oscillation is $\Delta n_*$, and we assume
$\Delta n_{*}\ll\bar{n}_{*}$. 
The dissipated energy per volume due to the oscillation is given by
\begin{equation}
\frac{d\epsilon}{dt}=-\zeta\left(\vec{\nabla}\cdot\vec{v}\right)^{2}\end{equation}
where $\vec v$ is the local velocity of the fluid of the conserved quantity and the continuity equation for its particle number $n_*$ reads
\begin{equation}
\frac{\partial n_{*}}{\partial t}+\vec{\nabla}\cdot\left(n_{*}\vec{v}\right)=0\,.\end{equation}
In the case that density varies slowly enough so that density gradients
can be neglected, and using $\Delta n_{*}/\bar{n}_{*}\ll1$, averaging
over a whole oscillation period $\tau=2\pi/\omega$ gives the bulk
viscosity as
\begin{equation}
\zeta\approx-\frac{2}{\omega^{2}}\left\langle \frac{d\epsilon}{dt}\right\rangle \frac{\bar{n}_{*}^{2}}{\left(\Delta\! n_{*}\right)^{2}}\,.
\label{eq:zetadef}
\end{equation}

Using the relationship between 
fluctuations in volume and fluctuations of a conserved quantity,
\begin{equation}
\frac{dn_{*}}{n_{*}}=-\frac{dV}{V}\label{eq:volume-fluctuation}\end{equation}
and the mechanical work done by a volume change
\begin{equation}
d\epsilon=-\frac{p}{V}dV\end{equation}
we can express
the dissipated energy per volume averaged over one time period
in terms of the induced pressure oscillation
\begin{equation}
\left\langle \frac{d\epsilon}{dt}\right\rangle=\frac{1}{\tau}\int_{0}^{\tau}\frac{p}{n_{*}}\frac{dn_{*}}{dt}dt\,.
\label{eq:dedt}
\end{equation}

To calculate the bulk viscosity we must calculate $p(t)$. We will assume that
the bulk viscosity arises from beta-equilibration of fermionic
species. We further assume that, in the absence of weak interactions, 
there would
be $s$ conserved species, and that there is a single channel of weak
interactions that can perform interconversion of species, leaving
$s-1$ exactly conserved fermion-number charges\footnote{In this counting we exclude fermions 
like neutrinos, which
escape from compact stars and so are effectively not conserved.}.
We defer discussion of the general situation of several coupled channels 
to future studies. Subtracting the chemical potentials of the
final state particles in the relevant weak channel from those of the
initial state particles, we obtain the difference
\begin{equation}
\mu_\Delta \equiv \sum_i \mu_i - \sum_f \mu_f \ .
\label{eq:delta-mu-def}
\end{equation}
which is the quantity that is driven out of equilibrium by the
driving density fluctuation, and whose re-equilibration
leads to bulk viscosity.
The quasi-equilibrium state can generally be described in terms
of the driving density $n_{*}$ and the ratio $x\equiv n_{1}/n_{*}$ 
where $n_1$ is the density of one of the particle species 
whose number is changed by the equilibration process.
For small oscillation amplitudes $\Delta n_{*}/\bar{n}_{*}\ll 1$
the pressure can then be expanded around its equilibrium value $\bar{p}=p\left(\bar{n}_{*}\right)$
\begin{equation}
p=\bar{p}
 + \left. \frac{\partial p}{\partial n_{*}}\right|_x \delta n_{*}
 + \left. \frac{\partial p}{\partial x}\right|_{n_*} \delta x\,,
\label{eq:pchange}
\end{equation}
where $\delta x$ is the deviation of $x$ from its beta-equilibrium value.
The $t$-independent part $\bar{p}$ as well as the term proportional
to the driving density fluctuation $\delta n_{*}$ do not contribute
to the viscosity integral. The remaining susceptibility can be 
rewritten
\begin{align}
\left.\frac{\partial p}{\partial x}\right|_{n_*}
 & = \bar{n}_{*}^{2}
 \left.\frac{\partial \mu_\Delta}{\partial n_{*}}\right|_x 
  \,.
\label{eq:susceptibility}
\end{align}
Because of weak interactions, $x$ depends on time,
\begin{equation}
\delta x(t)=\int_{0}^{t}\frac{dx}{dt^\prime}dt^\prime \ .
\label{eq:x}
\end{equation}
From Eqs.~\eqn{eq:zetadef},\eqn{eq:dedt},\eqn{eq:pchange},\eqn{eq:susceptibility},\eqn{eq:x},
\begin{equation}
\zeta=-\frac{1}{\pi}\frac{\bar{n}_{*}^{3}}{\Delta\! n_{*}}\int_{0}^{\tau}\frac{\partial \mu_\Delta}{\partial n_{*}}\int_{0}^{t}\frac{dx}{dt^\prime}dt^{\prime}\cos\left(\omega t\right)dt\,.
\end{equation}
We want to point out already at this point that, in contrast to the harmonic 
driving density oscillation $\delta n_{*}$ with amplitude $\Delta n_{*}$, the induced chemical potential fluctuation $\delta \mu_\Delta$ around the vanishing equilibrium value can have a more complicated anharmonic form. \\
The fluctuations of the density ratio can be obtained from an analogous
expansion of the chemical potential fluctuation
\begin{equation}
\delta\mu_\Delta=
\left.\frac{\partial \mu_\Delta}{\partial n_{*}} \right|_x \delta n_{*}
+\left.\frac{\partial \mu_\Delta}{\partial x}\right|_{n_*} \delta x
\end{equation}
which yields a linear equation relating $\mu_\Delta$ and $\delta x$
\begin{align}
\frac{d\mu_\Delta}{dt} & =C \omega\frac{\Delta n_{*}}{\bar{n}_{*}}\cos\left(\omega t\right)+B \bar{n}_{*}\frac{dx}{dt}\,,
\label{eq:mudot}
\end{align}
with the susceptibilities
\begin{align}
C\equiv\bar{n}_{*}\left.\frac{\partial\mu_\Delta}{\partial n_{*}}\right|_x
\quad,\quad 
B \equiv\frac{1}{\bar{n}_{*}}\left.\frac{\partial\mu_\Delta}{\partial x}\right|_{n_*}\label{eq:susceptibilities}\end{align}
Using \eqn{eq:mudot} we obtain
the bulk viscosity in terms of
the chemical potential fluctuation,
\begin{equation}
\zeta=-\frac{1}{\pi}\frac{\bar{n}_{*}}{\Delta\! n_{*}} \frac{C}{B} \int_{0}^{\tau}\mu_\Delta(t)\cos(\omega t)dt\,.
\label{eq:general-viscosity}
\end{equation}
In terms of a Fourier expansion of the periodic chemical potential fluctuation
\begin{equation}
\mu_\Delta(t)=\sum_{n=1}^{\infty}\left(a_{n}\sin\left( n \omega t\right)
+b_{n}\cos( n \omega t)\right)
\label{eq:trig-fourier-expansion}
\end{equation}
we see that the only component of $\mu_\Delta(t)$ that contributes to the 
viscosity is the component of the fundamental Fourier mode that lags
the driving volume oscillation by a phase of $\pi/2$.
This suggests that a truncated Fourier ansatz 
may provide a reliable approximation
for the viscosity: 
we will explore this idea in Sec.~\ref{sub:Strange-quark-matter}.

To obtain the temperature and amplitude dependence of the bulk viscosity,
we now discuss the general form of the beta equilibration rate.
We define the net equilibration rate
\begin{equation}
\Gamma^{(\leftrightarrow)} 
 \equiv \Gamma^{(\rightarrow)} -\Gamma^{(\leftarrow)}
 = \bar n_{*}\frac{\partial x}{\partial t} \,,
\end{equation}
where we use the convention that $\Gamma^{(\rightarrow)}$ is the rate for the
process where $n_1$ is decreased, and $\Gamma^{(\leftarrow)}$ is the rate
for the inverse process.
We study equilibration processes where the net rate takes the general form
\begin{equation}
\Gamma^{(\leftrightarrow)} = -\tilde{\Gamma} T^{\kappa}\mu_\Delta \left(1+\sum_{j=1}^{N}\chi_j\left(\frac{\mu_\Delta^{2}}{T^{2}}\right)^{j}\right)\,.
\label{eq:gamma-parametrization}\end{equation}
where $N$ is the highest power of $\mu_\Delta$
arising in the rate.
In terms of dimensionless variables
\begin{equation}
\varphi\equiv\omega t\quad,\quad {\cal A} \left(\varphi\right)\equiv\frac{\mu_\Delta\left(t\right)}{T}\end{equation}
the differential equation for the chemical fluctuation eq.~\eqn{eq:mudot} can be written as
\begin{equation}
\frac{d {\cal A}}{d\varphi}=d \cos\left(\varphi\right)-f {\cal A} \left(1+\sum_{j=1}^{N}\chi_{j}{\cal A}^{2j}\right)\,,
\label{eq:general-diff-eq}
\end{equation}
with the prefactors of the driving and feedback term given by
\begin{equation}
d\equiv\frac{C}{T}\frac{\Delta n_{*}}{\bar{n}_{*}}\quad,\quad f\equiv\frac{B\tilde{\Gamma}T^{\kappa}}{\omega}\,.
\label{eq:general-parameters}
\end{equation}
Note that the feedback term involves both linear and non-linear parts which are controlled by a single parameter $f$ and that its particular form is determined by the constants $\chi_j$ which parametrize the particular weak rate. The viscosity is then finally given by
\begin{align}
\zeta & =\frac{T C}{\pi\omega B}\frac{\bar{n}_{*}}{\Delta n_{*}}\int_{0}^{2\pi}{\cal A}\left(\varphi;d,f\right)\cos\left(\varphi\right)d\varphi\,,\label{eq:general-solution}\end{align}
where ${\cal A}$ is the periodic solution to eq. \eqn{eq:general-diff-eq}.\\
Before we discuss the general solution of these equations in detail,
let us consider its asymptotic limits. 

\noindent \underline{Sub-thermal limit}\\
In the limit $\mu_\Delta \ll T$ corresponding to ${\cal A} \ll1$ the non-linear
terms can be neglected
\begin{equation}
\left( \frac{d}{d\varphi}-f \right){\cal A} =d \cos\left(\varphi\right)\,.\end{equation}
Since this equation is linear, the fluctuation ${\cal A}$ must
be harmonic and only the $n=1$ term in the Fourier ansatz eq.~(\ref{eq:trig-fourier-expansion})
is present. Inserting this ansatz yields the solution for the required
Fourier coefficient
\begin{equation}
b_1=-\frac{d f}{1+f^2}\,.\end{equation}
Inserted in eq.~(\ref{eq:general-solution}) 
this yields the general sub-thermal result, denoted
by a superscript $<$, for the bulk viscosity of an arbitrary form of matter which shows the characteristic resonant form 
\begin{equation}
\zeta^{<}=\frac{C^{2}\tilde{\Gamma}T^{\kappa}}{\omega^{2}+(B \tilde{\Gamma} T^{\kappa})^2}
= \zeta_{max}^< \frac{2 \omega B\tilde\Gamma T^\kappa}{\omega^2 
+ (B\tilde\Gamma T^\kappa)^2}
\,.
\label{eq:sub-viscosity}
\end{equation}
As long as the combination of susceptibilities $C^2/B$ does not
vary too quickly with temperature, the sub-thermal viscosity has a maximum
\begin{equation}
\zeta_{max}^{<}=\frac{C^{2}}{2\omega B}\quad\mathrm{at}\quad T_{max}=\left(\frac{\omega}{\tilde{\Gamma}B}\right)^{\frac{1}{\kappa}}\,.
\label{eq:sub-maximum}
\end{equation}

\noindent \underline{Supra-Thermal limit}\\
The opposite, suprathermal limit, $\mu_\Delta\gg T$, corresponds to ${\cal A} \gg1$.
Since the feedback term in the
differential equation is restraining, this limit can only be reached
in the limit of large driving terms $d \gg1$.
In this case only the largest power of ${\cal A}$ is relevant and
eq.~(\ref{eq:general-diff-eq}) reduces to
\begin{equation}
0=d\cos\left(\varphi\right)-\chi_{N}f {\cal A}^{2N+1}\;\Rightarrow\;{\cal A}\sim\left(\frac{\Delta n_{*}}{\bar{n}_{*}}\right)^{\frac{1}{2N+1}}\,.
\end{equation}
The viscosity scales correspondingly in this limit as
\begin{equation}
\zeta\sim\left(\frac{\Delta n_{*}}{\bar{n}_{*}}\right)^{-\frac{2N}{2N+1}}\label{eq:supra-limit}\end{equation}
and decreases at very
large amplitudes.

\begin{figure*}
\begin{minipage}[t]{0.5\textwidth}%
\begin{center}\underline{$d=1$}\end{center}
\includegraphics[scale=0.8]{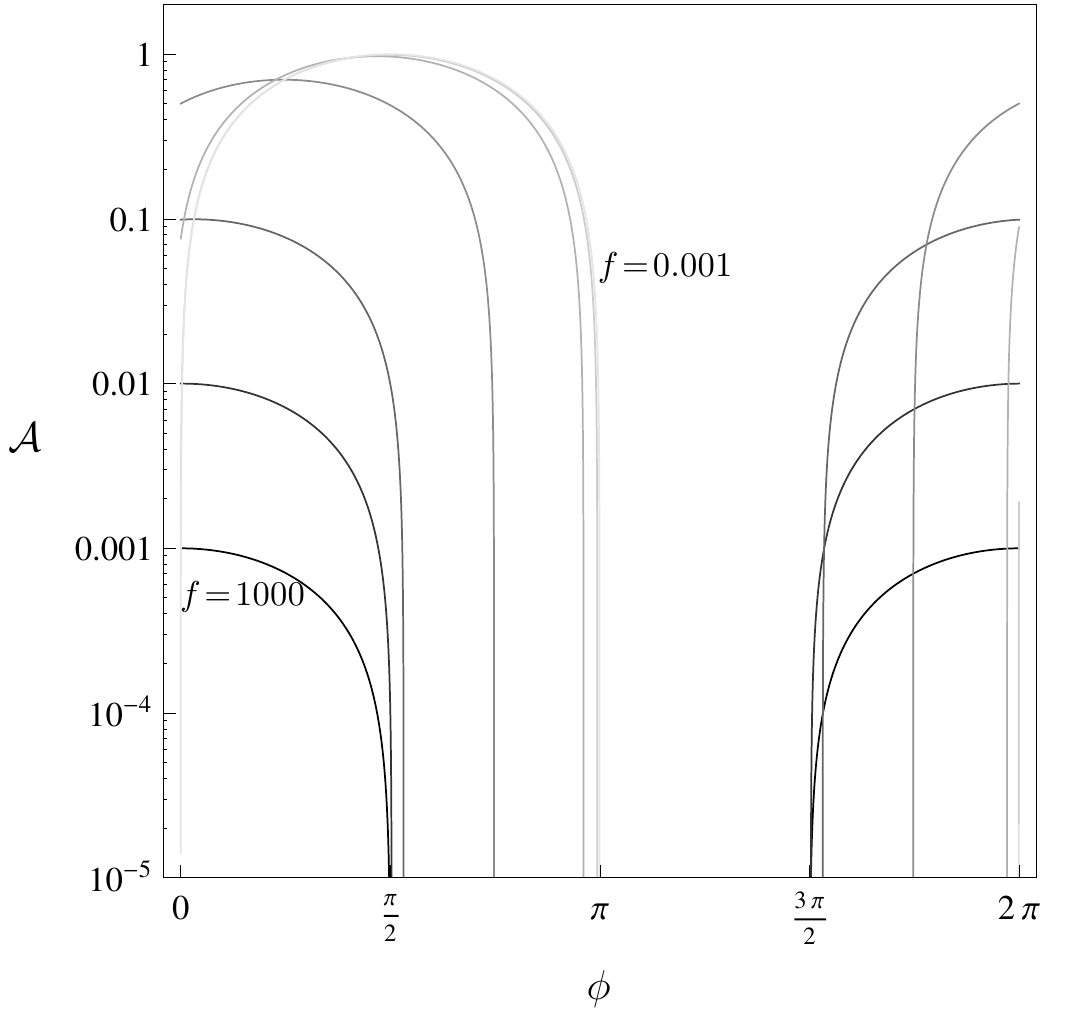}%
\end{minipage}%
\begin{minipage}[t]{0.5\textwidth}%
\begin{center}\underline{$f=1$}\end{center}
\includegraphics[scale=0.8]{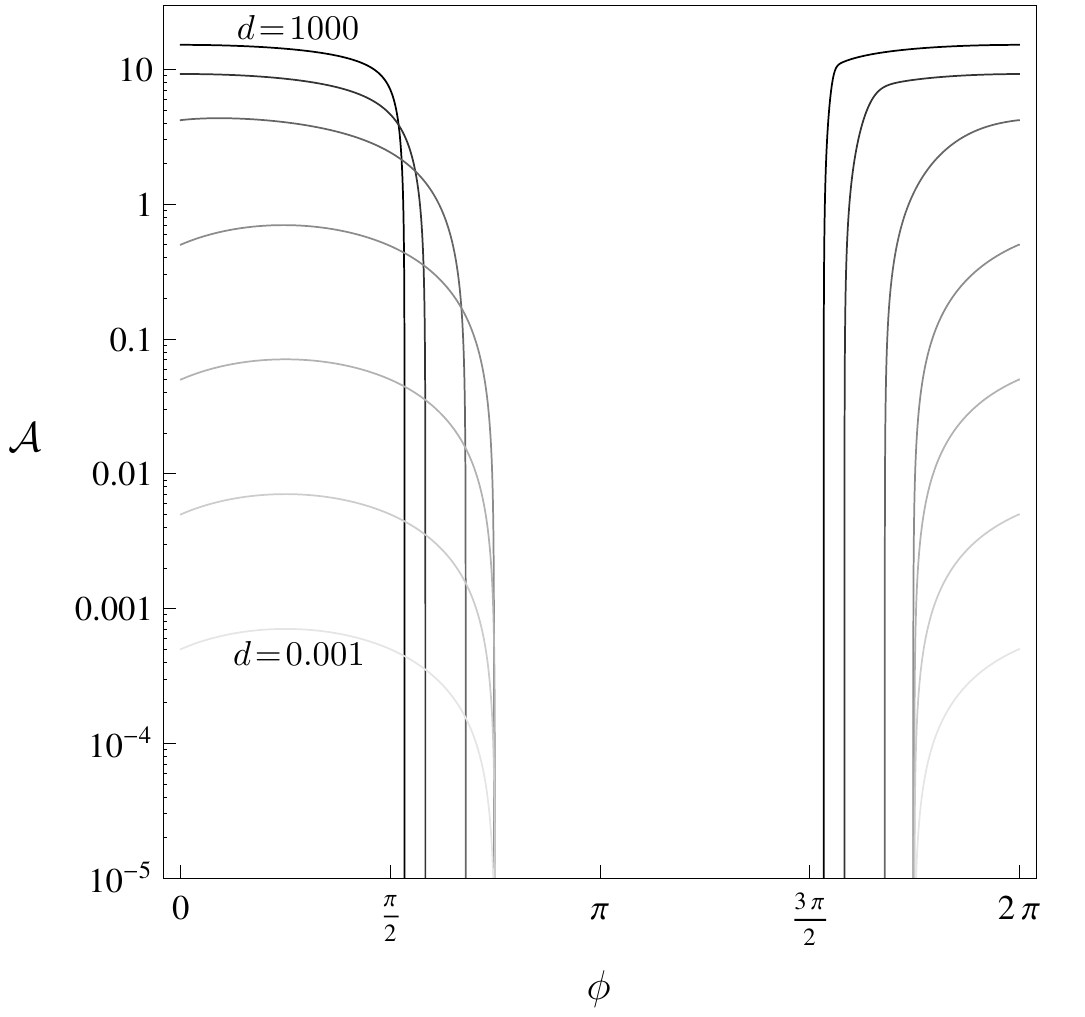}%
\end{minipage}

\caption{\label{fig:waveforms} 
Waveform ${\cal A}\left(\phi\right)=\mu_\Delta\left(\omega t\right)/T$
for different values of the two independent parameters. We show only
the positive half-wave, on a logarithmic scale. \emph{Left panel:} 
Fixed driving term $d=1$, with varying feedback term
$f=0.001,0.01,\cdots,1000$. As $f$ rises, the phase lag increases from
zero towards $\pi/2$, but at the same time the amplitude decreases as $1/f$.
\emph{Right panel:} Fixed feedback term $f=1$, with varying
driving term $d=0.001,0.01,\cdots,1000$. As $d$ rises,
the phase lag rises from $\pi/4$ to $\pi/2$ and the waveform
becomes increasingly anharmonic, approaching a square wave in the limit.
}
\end{figure*}

\noindent \underline{General solution}\\
After these limiting cases we will discuss the qualitative aspects
of the general solution eq.~(\ref{eq:general-solution}). 
Due to the non-linearity of the differential equation (\ref{eq:general-diff-eq}) this requires
a numeric solution. Yet, for each weak channel, characterized by the constants $\chi_{j}$, such a solution as a function of the two independent variables $d$ and $f$ has to be performed only once and is then valid for any equation of state and includes the complete dependence on the underlying parameters in eq.~\eqn{eq:general-parameters}.
The qualitative behavior of the solution as a function of the two
independent parameters $d$ and $f$ is shown for 
hadronic matter with modified Urca process in fig.~\ref{fig:waveforms}.
Turning up the feedback term at fixed driving term increases the phase
shift of the waveform from $0$ to $\pi/2$ and at the same time decreases
the amplitude, but the waveform stays harmonic. In contrast, turning
up the driving term at fixed feedback increases the amplitude towards
the supra-thermal regime ${\cal A}>1$ and the waveform becomes increasingly
anharmonic. Recall, however, that only the phase shifted harmonic
component in the Fourier expansion contributes to the viscosity eq.
(\ref{eq:general-solution}).

Motivated by the above expression eq.~(\ref{eq:sub-maximum}) for
the maximum in the sub-thermal regime the general result can be written
in the form
\begin{align}
\zeta & =\zeta^<_{max} \, {\cal I} \left(d,f\right)= \frac{C^2}{2\omega B}{\cal I} \left(d,f\right) \label{eq:general-parametrization}\end{align}
where the dimensionless function ${\cal I}$ that includes the
non-trivial parameter dependence is given by

\begin{equation}
{\cal I}(d,f)\equiv\frac{2}{\pi d}\int_{0}^{2\pi} {\cal A}(\varphi;d,f)\cos(\varphi)d\varphi\end{equation}
The expression ${\cal I}$ can then
be tabulated as a function of the independent parameters $d$ and $f$.
We believe that presenting our results in this form will make them
easier to apply to calculations of r-mode damping, where the complete parameter dependence is required.
The computation of the damping time
of the mode involves an integral over the star of an expression that
involves the bulk viscosity (e.g., \cite{Owen:1998xg}) which varies throughout the star because of its dependence on the amplitude of the mode and the susceptibilities, both of which are position-dependent.
The function ${\cal I}(d,f)$ encapsulates the dependence of the bulk viscosity on the position-dependent parameters, allowing straightforward evaluation of the damping time integral.

\begin{figure*}
\begin{minipage}[t]{0.5\textwidth}%
\begin{center}(a)~\underline{Hadronic matter, 
  modified Urca process}\end{center}
\includegraphics[width=\hsize]{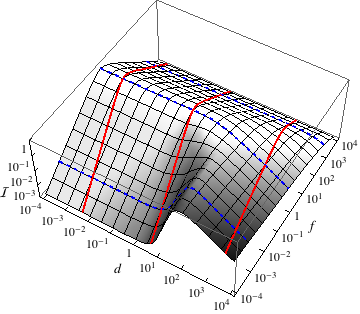}%
\end{minipage}%
\begin{minipage}[t]{0.5\textwidth}%
\begin{center}(b)~\underline{Quark matter, 
  non-leptonic process}\end{center}
\includegraphics[width=\hsize]{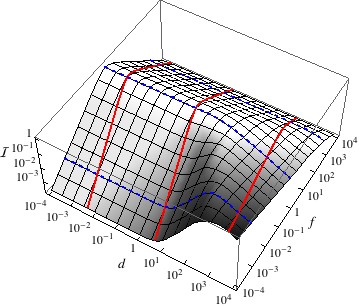}%
\end{minipage}

\caption{\label{fig:gen-sol} The function ${\cal I}$ arising in the general
solution eq.~(\ref{eq:general-solution}) for two models of dense matter.
Left panel: hadronic matter with modified Urca equilibration.
Right panel: quark matter with the non-leptonic equilibration process 
eq.~\eqn{eq:nonleptonic}.
The function has a global maximum of $1$
reached asymptotically for $d\to 0,f=1$ and a line of slowly decreasing
local maxima along a parabola in the $d$-$f$ plane. The shading
of the surface denotes the size of the amplitude ${\cal A}$ so that
dark shades of grey represent the supra-thermal regime. 
Eq.~(\ref{eq:general-parameters}) relates $d$ and $f$ to underlying
physical parameters such as temperature $T$ and amplitude.
An amplitude increase (keeping all other variables fixed) results in a linear increase in the variable
$d$ as shown by the dashed (blue online) curves. An increase in temperature
changes the viscosity along a line
shown by the solid (red online) curves. 
}

\end{figure*}

The function ${\cal I}(d,f)$ is shown in fig.~\ref{fig:gen-sol} for two
examples: a model of strange quark matter and 
a model of hadronic matter; details of the models are discussed
below. We see that the function has
the same qualitative form in both cases.
It has a global maximum value of 1, reached in the
sub-thermal limit
and a line of local maxima along a parabola in the $d$-$f$-plane.
Thus the maximum value (\ref{eq:sub-maximum}) of the sub-thermal viscosity
is also the maximum in the general case and depends only
on the equation of state, the density and the frequency but is independent
of the weak rate. The weak rate influences, however, at what temperatures
and amplitudes the local maxima are reached. 
As seen from eq.~(\ref{eq:general-parameters}),
the parameter $d$ is directly proportional to the amplitude, so that
at moderate feedback an amplitude increase does initially not affect
the viscosity at all, corresponding to the amplitude-independent sub-thermal
result. But once the amplitude becomes sufficiently large we enter the
supra-thermal regime and the viscosity
increases strongly by orders of magnitude until it reaches its maximum. 
The size of the amplitude
${\cal A}$ is denoted in fig.~\ref{fig:gen-sol} by the darkness of shading 
of the surface. This qualitative behavior has already been
observed in \cite{Madsen:1992sx} but we find that at even higher
amplitudes the viscosity decreases again according to the limiting
behavior eq.~(\ref{eq:supra-limit}). In contrast, at large feedback
the viscosity becomes basically amplitude independent over the relevant
parameter range as described by the sub-thermal result.

\begin{figure}
\includegraphics[width=\hsize]{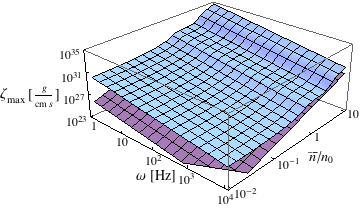}
\caption{\label{fig:max-viscosity}The maximum viscosity in the supra-thermal
limit eq.~(\ref{eq:sup-maximum}) of hadronic matter (upper surface) and a hadronic gas (lower surface) as a function
of baryon density and angular frequency. This represents also the analytic
prefactor of the general expression in eq.~(\ref{eq:general-solution}).
The corresponding plot for strange quark matter eq.~(\ref{eq:strange-maximum})
would be trivial since it does not depend on the density to leading
order and only shows the analytic $1/\omega$ dependence.}
\end{figure}

Let us now discuss the dependence of the viscosity on the underlying parameters in eq.~\eqn{eq:general-parameters}. An amplitude increase (keeping all other variables fixed) results in a linear increase in the variable $d$ as shown by the dashed (blue online) curves in fig.~\ref{fig:gen-sol}. An increase in temperature
changes the viscosity along a line shown by the solid (red online) curves. In order to assess the frequency and amplitude dependence we must take into account the prefactor in eq.~\eqn{eq:general-parametrization}.
This prefactor, given by the maximum viscosity in the subthermal regime, is shown in fig.~\ref{fig:max-viscosity}, for the hadronic model of fig.~\ref{fig:gen-sol}(a). It exhibits a monotonic increase with density and inverse angular frequency.
An increase in angular frequency therefore changes the viscosity via a change of ${\cal I}$ towards the negative $f$-direction and furthermore via the overall prefactor featuring an additional $1/\omega$
dependence. A density increase has an even more indirect impact since
it depends on the detailed form of the susceptibilities $C\left(\bar{n}_{*}\right)$
and $B\left(\bar{n}_{*}\right)$ which likewise arise in the prefactors
of the viscosity. These dependencies will be studied in more detail below.

\section{Strange quark matter\label{sub:Strange-quark-matter}}

\subsection{General features}

It has been suggested that the true ground state of matter at
high densities may be strange quark matter
\cite{Bodmer:1971we,Witten:1984rs} consisting of $u$, $d$ and $s$
quarks. In that case self-bound strange stars could
exist. In this section we apply the results obtained above to
strange quark matter, which is both an interesting physical
scenario and a useful introductory example in which one can make
illuminating simplifications which are not possible for
the case of hadronic matter discussed in the next section. The bulk
viscosities of various forms of strange quark matter have previously
been analyzed
\cite{Madsen:1992sx,Manuel:2004iv,Alford:2006gy,Sa'd:2006qv,Alford:2007pj,Alford:2007rw,Manuel:2007pz,Sa'd:2007ud,Dong:2007ax}, and
the influence of strong magnetic fields
has recently been discussed \cite{Huang:2009ue}.

The dominant channel in unpaired strange quark matter is the non-leptonic
flavor changing process
\begin{equation}
d+u\leftrightarrow s+u \ .
\label{eq:nonleptonic}
\end{equation}
The corresponding quark Urca processes, which involve leptons,
are parametrically suppressed in the ratio $T/\mu_q$. 
The conserved quantity that tracks the driving oscillation can be
chosen as the baryon number,
with density $n=\tfrac{1}{3}\left(n_{s}+n_{d}+n_{u}\right)$.
The equilibrating chemical potential eq.~\eqn{eq:delta-mu-def} carries in this case the quantum numbers of neutral $K$-mesons and is therefore denoted by
\begin{equation}
\mu_{K}\equiv\mu_{s}-\mu_{d}\end{equation}
The rate of the non-leptonic process eq. \eqn{eq:nonleptonic} is
given by \cite{Heiselberg:1992bd}
\begin{equation}
\Gamma^{(\leftrightarrow)}_{q}=-\frac{16}{5\pi^{5}}G_{F}^{2}
  \sin^{2}\!\theta_{C}\cos^{2}\!\theta_{C}\mu_{d}^{5}\mu_K
  \left(4\pi^{2}T^{2}+\mu_K^{2}\right)
\end{equation}
From this expression one can directly
obtain the equilibration rate parameters
$\tilde{\Gamma}$, $\kappa$ and $\chi_{i}$ in the parameterization
eq.~(\ref{eq:gamma-parametrization}). Their values are given
in the first row of table~\ref{tab:weak-parameters}.

\begin{table*}[htb]
\newcommand{\st}{\rule[-3ex]{0em}{8ex}}  
\begin{tabular}{l@{\quad}c@{\quad}c@{\quad}c@{\quad}c@{\quad}c}
\hline 
\rule[-1.5ex]{0em}{4ex} Matter/Channel
  & $\tilde{\Gamma} \,\bigl[\MeV^{(3-\kappa)}\bigr]$ 
  & $\kappa$ & $\chi_{1}$ & $\chi_{2}$ & $\chi_{3}$\tabularnewline
\hline 
\st quark non-leptonic & $\dsp 6.59\!\times\!10^{-12}\, \Bigl(\frac{\mu_{d}}{300~\MeV}\Bigr)^{5}$ 
  & $2$ & $\dsp\frac{1}{4\pi^{2}}$ & $0$ & $0$\tabularnewline
\hline 
\st hadronic direct Urca & $\dsp 5.24\!\cdot\!10^{-15} \left(\!\frac{x\, n}{n_{0}}\!\right)^{\!\frac{1}{3}}$ 
  & $4$ & $\dsp\frac{10}{17\pi^{2}}$ & $\dsp\frac{1}{17\pi^{4}}$ & $0$\tabularnewline
\hline 
\st hadronic modified Urca & $\dsp 4.68\!\cdot\!10^{-19} \left(\!\frac{x\, n}{n_{0}}\!\right)^{\!\frac{1}{3}}$ 
  & $6$ & $\dsp\frac{189}{367\pi^{2}}$ & $\dsp\frac{21}{367\pi^{4}}$ & $\dsp\frac{3}{1835\pi^{6}}$\tabularnewline
\hline
\end{tabular}
\caption{\label{tab:weak-parameters}Weak interaction parameters describing
the considered damping process. Here $\mu_q$ is the quark chemical potential, $n$ is the baryon density, $n_0$ nuclear saturation density and $x$ the proton fraction.
}
\end{table*}

\subsection{Analytic approximation}

Since in this case only cubic non-linearities arise it is possible
to obtain an approximate analytic solution to the non-linear equation
(\ref{eq:general-diff-eq}). Taking into account the above observation
that only the leading Fourier coefficient in the expansion of the
chemical potential oscillation contributes to the bulk viscosity it
is natural to seek such a solution via a Fourier ansatz up to a given
order $O$
\begin{equation}
{\cal A}\left(t\right)=\sum_{n=-O}^{O}\tilde{\cal A}_{n}\mathrm{e}^{in\omega t}\end{equation}
where the complex form is used to simplify the computation. 
In principle, the amplitude of the 
leading Fourier mode will depend on the truncation order $O$,
but analytically solving eq.~(\ref{eq:general-diff-eq})
via a computer algebra system to order $O=2$ we find that the coefficients $\tilde{\cal A}_{\pm 2}$ vanish identically. Correspondingly
anharmonicities do not directly contribute to the viscosity
and are even absent to next to leading order so that we can restrict
our analysis to the leading order $O=1$. Although such a parameterization
neglects any anharmonicities it properly captures both the amplitude
and the phase shift of the oscillation even in the large amplitude
regime. Due to the reality of the solution there is only one independent
complex Fourier exponent determined by a non-linear algebraic equation.
In the case of quark matter where $\chi_{i}=0$ for $i>1$ and only
the leading non-linear term $\chi\left(\mu_K /T\right)^{3}$ is
present an analytic solution of this equation is possible. In this
case we can decompose the amplitude into real and imaginary parts
$\tilde{\cal A}_{1} = A_R + i A_I$, obeying coupled equations
\begin{align}
 f A_R \bigl( 1 + 3\chi (A_R^2 + A_I^2)\bigr) + A_I &= -d/2 \\
 f A_I \bigl( 1 + 3\chi (A_R^2 + A_I^2)\bigr) - A_R &= 0
\end{align}
Note that an analytic solution is only possible
because the quark matter equations are cubic; other forms of matter with higher
order non-linearities in eq.~(\ref{eq:gamma-parametrization}) require
a numeric solution. The above system of algebraic equations has a
lengthy analytic solution which we refrain from giving here
because, as we will see below, it can be very accurately approximated
by a much simpler expression \eqn{eq:viscosity-parametrization}
constructed from a combination of the solutions
in the sub-thermal and supra-thermal regimes. Therefore
we now concentrate on the supra-thermal case, denoted by the index $>$,
where the temperature-dependent term can be neglected,
\begin{align}
&{\cal A}^{>}(\varphi)=2 A_R\cos\left(\varphi\right)-2 A_I\sin\left(\varphi\right)\nonumber \\
&=-\frac{3d}{2}\left(\frac{\left(q(z)^2-1\right)^{2}}{\sqrt{3} z\,q(z)^2}\cos(\varphi)
 +\frac{q(z)^2-1}{z\,q(z)}\sin(\varphi)\right)\,,
\end{align}
where the dimensionless quantity $z$ is defined by
\begin{equation}
z\equiv\frac{9\sqrt{3}}{8}\chi d^{2}f=\frac{9\sqrt{3}\chi}{8}\frac{\tilde{\Gamma}C^{2}B}{\omega}T^{\kappa-2}\left(\frac{\Delta\! n_{*}}{\bar{n}_{*}}\right)^{2}\end{equation}
and
\begin{equation}
q(z) \equiv \Bigl(\sqrt{z^{2}+1}-z\Bigr)^{\frac{1}{3}} \ .
\end{equation}
Eq.~(\ref{eq:general-solution}) then yields the approximate analytic
result for the bulk viscosity in the supra-thermal regime
\begin{equation}
\zeta^{>}\approx\frac{2}{3\sqrt{3}}\frac{C^{2}}{B\omega}
 \, h\biggl(\frac{9\sqrt{3}\chi}{8}\frac{\tilde{\Gamma}C^{2}B}{\omega}T^{\kappa-2}\left(\frac{\Delta\! n_{*}}{\bar{n}_{*}}\right)^{2}\biggr)
\label{eq:supra-viscosity}\end{equation}
in terms of the dimensionless function
\begin{equation}
h(z)=\frac{9}{4z}\left(\left(\sqrt{z^{2}\!+\!1}\!-\! z\right)^{\frac{2}{3}}\!+\!\left(\sqrt{z^{2}\!+\!1}\!+\! z\right)^{\frac{2}{3}}\!-\!2\right)\,.\label{eq:h-function}\end{equation}
This function has a maximum at $z_{max}=3\sqrt{3}$.
Since $h(z_{max})=3\sqrt{3}/4$,
the corresponding maximum value of the viscosity is
\begin{equation}
\zeta_{max}^{>}=\frac{2}{3\sqrt{3}}\frac{C^{2}}{B\omega}
 \, h(z_{max})=\frac{C^{2}}{2B\omega}\label{eq:sup-maximum}\end{equation}
which strikingly is the same expression as in the
sub-thermal limit eq.~(\ref{eq:sub-maximum}). Correspondingly the
bulk viscosity has a universal upper bound $\zeta_{max}$ that is
\emph{independent} of the particular weak damping process. It is directly
proportional to the oscillation period with a coefficient that only
depends on the response of the strongly interacting matter. However,
the corresponding temperature (\ref{eq:sub-maximum}) and amplitude
\begin{equation}
\left(\frac{\Delta n_{*}}{\bar{n}_{*}}\right)_{max}=\sqrt{\frac{8\omega}{3\chi\tilde{\Gamma}T^{\kappa-2}C^{2}B}}\end{equation}
at which this maximum is reached both depend on the weak rate.

Knowing the upper bound $\zeta_{max}$ and the
functional behavior in the extreme sub-thermal and supra-thermal
limits, allows us to give a simple parameterization of the full function
for all temperatures and amplitudes.
We construct a weighted sum of the analytic results in the sub-thermal eq.~(\ref{eq:sub-viscosity}) and the supra-thermal regime eq.~(\ref{eq:supra-viscosity}),

\begin{widetext}
\begin{align}
\zeta_{par} & \approx\zeta^{<}+\theta(T_{max}-T)\frac{\zeta_{max}-\zeta^{<}}{\zeta_{max}}\zeta^{>}\label{eq:viscosity-parametrization}\\
 & =\frac{C^{2}}{2B\omega}\left(\frac{2\omega\tilde{\Gamma}BT^{^{\kappa}}}{\omega^{2}+\tilde{\Gamma}^{2}B^{2}T^{^{2\kappa}}}
 +\theta\Biggl(\biggl(\frac{\omega}{\tilde{\Gamma B}}\biggr)^{\frac{1}{\kappa}}
   -T\Biggr)
  \frac{4}{3\sqrt{3}}\frac{\left(\omega-\tilde{\Gamma}BT^{^{\kappa}}\right)^{2}}{\omega^{2}+\tilde{\Gamma}^{2}B^{2}T^{^{2\kappa}}}
 h\Biggl(\frac{9\sqrt{3}\chi}{8}\frac{\tilde{\Gamma}BC^{2}T^{\kappa-2}}{\omega}\left(\frac{\Delta n_{*}}{\bar{n}_{*}}\right)^{2}\Biggr)\right)\nonumber 
\end{align}

\end{widetext}

Studies of the damping of compact star oscillations previously took
into account only the first, sub-thermal term in the parameterization
eq.~(\ref{eq:viscosity-parametrization}). The simple analytic form
allows one to conveniently extend these studies in order to include large
amplitude effects encoded in the second term. The small deviations
of the simplified parameterization eq.~(\ref{eq:viscosity-parametrization})
from the exact value of the bulk viscosity are
negligible compared to the considerable uncertainties inherent
in such a damping analysis. Evaluation of this expression requires
knowledge of the susceptibilities $B$ and $C$ that depend on the equation of
state. 

\subsection{Models of quark matter}

We now apply the results derived above to some simple models of quark matter.
We start with the simplest model, 
free quarks in a ``confining bag''.
We will call this a ``quark gas'' (QG).
We consider a 3-flavor quark and electron gas, with massless electron, up and down quarks and strange quark of mass $m_s$ with pressure
\begin{align}
p_{QG} & = \frac{1}{4\pi^{2}}\biggl(
  \mu_{d}^{4}\!+\!\mu_{u}^{4}\!+\!\mu_{s}p_{Fs}^3
  \!-\!\frac{3}{2}m_{s}^{2}\mu_{s}p_{Fs} \nonumber \\
&+\frac{3}{2}m_{s}^{4}\log\Bigl(\frac{\mu_{s}+p_{Fs}}{m_{s}}\Bigr)\biggr)-{\cal B}+\frac{\mu_{e}^{4}}{12\pi^{2}} 
\label{eq:quark-pressure}
\end{align}
where the strange quark Fermi momentum is given by $p_{Fs}^2=\mu_s^2-m_s^2$.
Here ${\cal B}$ is the phenomenological bag constant that is important for the 
equilibrium composition of a strange star, but does not affect transport properties like the bulk viscosity studied in this work. The equilibrium state is determined from eq.~(\ref{eq:quark-pressure}) by taking into account charge neutrality and weak equilibrium with respect to both the explicitly considered non-leptonic channel as well as the quark Urca channel. 

In quark matter there are  multiple channels for beta equilibration: 
as well as the  nonleptonic channel \eqn{eq:nonleptonic}
there are Urca channels which convert $d$ or $s$ quarks in to $u$ quarks
and electrons, and emit neutrinos. However, at 
temperatures and oscillation frequencies of interest for compact star 
physics the Urca rates are much slower, and their contribution to the
bulk viscosity is heavily suppressed. This means that the 
fractions $x_u$ and $x_e$ remain constant during the oscillation.
The required susceptibilities then are given by
\begin{align}
C_{q} & =\bar{n}\left(\frac{\partial\mu_{s}}{\partial n}-\frac{\partial\mu_{d}}{\partial n}\right)_{x_s,x_u,x_e} \,,\\
B_{q} & =\frac{1}{\bar{n}}\left(\frac{\partial\mu_{s}}{\partial x_{s}}-\frac{\partial\mu_{d}}{\partial x_{s}}\right)_{n,x_u,x_e}\,.\end{align}
Taking into account charge neutrality, the above equation of state yields to leading order in $m_s/\mu_q$ the susceptibilities given in table~\ref{tab:strong-parameters} (for the case $c=0$).

\begin{table*}[htb]
\newcommand{\st}{\rule[-3.5ex]{0em}{8.5ex}}  
\begin{tabular}{l@{\quad}c@{\quad}c@{\quad}c}
\hline 
\rule[-2ex]{0em}{5ex} & & $B$ & $C$\tabularnewline
\hline 
\st quark matter (gas: c=0) & & $\dsp\frac{2\pi^{2}}{3(1-c)\mu_{q}^{2}}\left(1\!+\!\frac{m_{s}^{2}}{12(1-c)\mu_{q}^{2}}\right)$ & $\dsp-\frac{m_{s}^{2}}{3(1-c)\mu_{q}}$\tabularnewline
\hline 
\st hadronic matter & & $\dsp\frac{8S}{n}\!+\negthinspace\frac{\pi^{2}}{\left(4\left(1\!-\!2x\right)S\right)^{2}}$ & $\dsp 4\!\left(1\!-\!2x\right)\!\left(\! n\!\frac{\partial S}{\partial n}\!-\!\frac{S}{3}\!\right)$\tabularnewline
\hline
\st free hadron gas & & $\dsp\frac{4m_{N}^{2}}{3\left(3\pi^{2}\right)^{\frac{1}{3}}n^{\frac{4}{3}}}$ & $\dsp\frac{\left(3\pi^{2}n\right)^{\frac{2}{3}}}{6m_{N}}$\tabularnewline
\hline
\end{tabular}
\caption{\label{tab:strong-parameters}
Strong interaction parameters describing
the response of various models of dense matter. In the case of 
hadronic matter with baryon density $n$ a quadratic ansatz in the proton fraction $x$ 
parameterized
by the symmetry energy $S$ eq.~\eqn{eq:quadratic}
is employed. The expressions for a free hadron
gas are given to leading order in $n/m_{N}^{3}$,
and for quark matter with quark chemical potential $\mu_q$ using eq.~\eqn{eq:quark-eos-parametrization}
to next to leading order in $m_{s}/\mu_q$. The parameter $c$ takes into account interaction effects within the employed quark matter model and vanishes for an ideal quark gas.
}
\end{table*}

According to eq.~(\ref{eq:sup-maximum}) the maximum viscosity
of a quark gas is given by
\begin{equation}
\zeta_{max}\approx\frac{m_{s}^{4}}{12\pi^{2}\omega}\label{eq:strange-maximum}\end{equation}
which depends on density only through possible
density-dependence of the strange quark mass.

\begin{figure}
\includegraphics[scale=0.8]{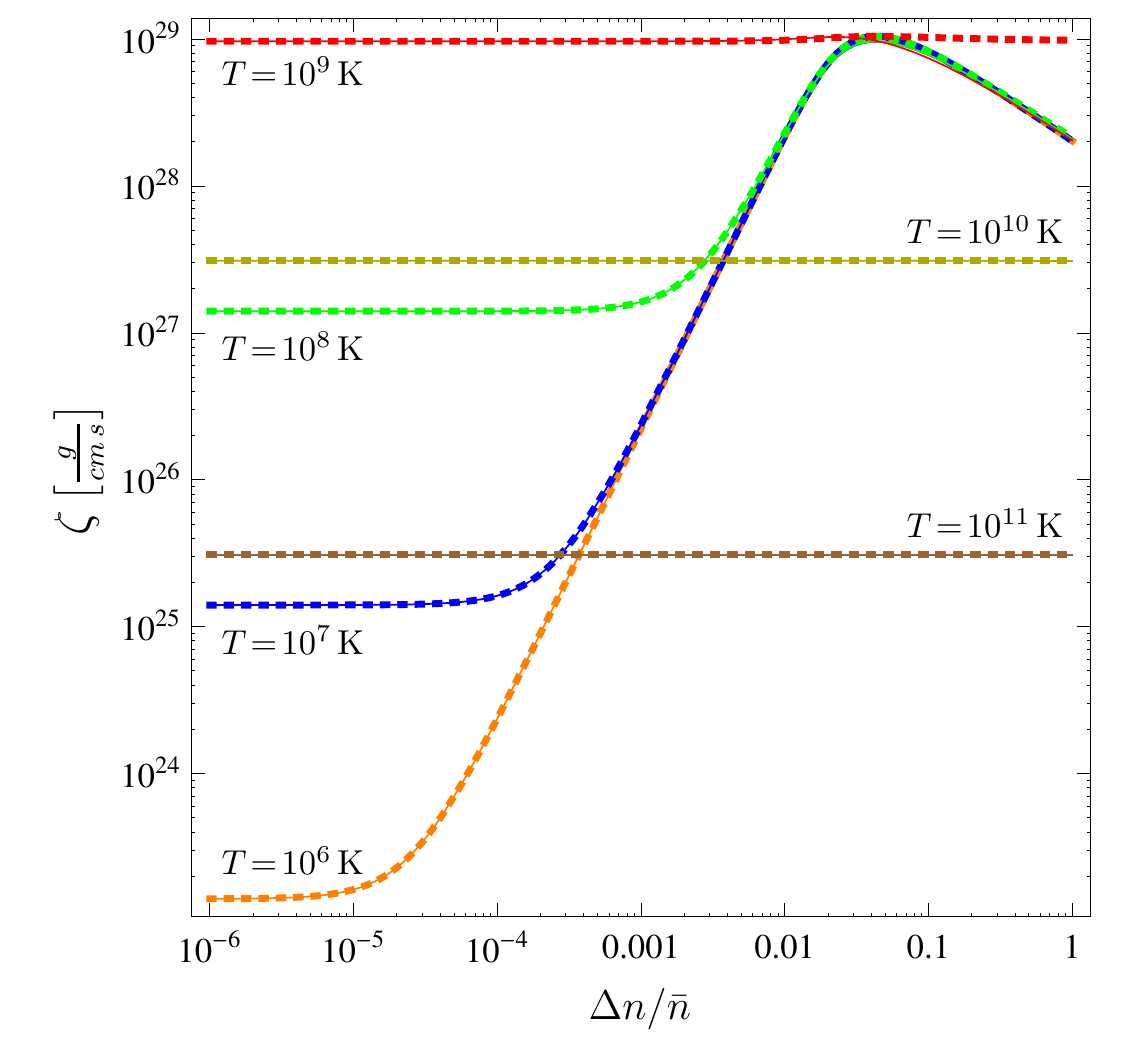}
\caption{\label{fig:quark-viscosity} The viscosity of a strange quark gas
as a function of the amplitude of the density oscillation $\Delta n/\bar n$
for different temperatures. The plots are given for an intermediate density $\bar n=2n_0$ and a frequency $\omega=8.4$ kHz corresponding to the oscillation frequency $\omega=4/3\,\Omega$ of the quadrupole r-mode of a millisecond pulsar. The viscosity increases with the given temperatures starting from $10^6$ K (bottom) to $10^9$ K
(top) and then decreases again. The thick, dashed curves represent the 
analytic model parametrization eq.~\eqn{eq:viscosity-parametrization}
and the thin, full curves beneath them give the
full numeric result. Clearly the parametrization is very accurate
in the relevant regime below the maximum. At high temperatures the viscosity does not reach the supra-thermal regime for any physical value of the amplitude, hence the
horizontal lines for $T\geqslant 10^{10}$\,K.
}
\end{figure}

In Fig.~\ref{fig:quark-viscosity} we give a comparison between the
parameterization eq. (\ref{eq:viscosity-parametrization}) 
and the full numeric solution. We show the amplitude dependence
of the viscosity of strange quark matter for a range of temperatures
These results are analogous to those given by Madsen in his initial analysis
of supra-thermal effects \cite{Madsen:1992sx}. The analytic solution
features the qualitative form that has been observed for the general
result in fig.~\ref{fig:gen-sol} and shows a striking agreement with
the full solution in the physically relevant region of amplitudes
below the maximum. Note that for temperatures around $T_{max}$ the parametrization eq.~(\ref{eq:viscosity-parametrization}) overestimates the viscosity for amplitudes above $\left(\Delta n/\bar n\right)_{max}$, as can be seen for the $T=10^9$ K curve in fig.~\ref{fig:quark-viscosity}. However, if such amplitudes are reached then suprathermal bulk viscosity
is overwhelmed, and other physics will have to be
invoked to stop the growth of the mode.

We now examine the sensitivity of our results to uncertainties
in the quark matter equation of state.
We use an extension of the phenomenological parameterization 
proposed in \cite{Alford:2004pf} that allows us to study the behavior of the equation of state around chemical equilibrium. Expanding the ideal gas pressure to quartic order in $m_s$, the $m_s$-independent quartic terms in the individual quark are modified
\begin{align}
p_{par}=&\frac{1-c}{4\pi^{2}}\left( \mu_{d}^{4}+\mu_{u}^{4}+\mu_{s}^{4} \right)-\frac{3 m_{s}^{2}\mu_{s}^{2}}{4 \pi^{2}}\nonumber \\ 
&+\frac{3}{32 \pi^2}\left( 3+4\log\left( \frac{2\mu_s}{m_s}\right)\right)-{\cal B}+\frac{\mu_{e}^{4}}{12\pi^{2}}
\label{eq:quark-eos-parametrization}
\end{align}
where $c$ is a new parameter which incorporates some effects of
strong interactions between the quarks and $m_s$ 
can parametrize here, in addition to corrections arising from the strange quark mass, also other interaction effects, like the pairing gap in color superconducting matter \cite{Alford:2004pf}.

\begin{figure}
\includegraphics[scale=0.8]{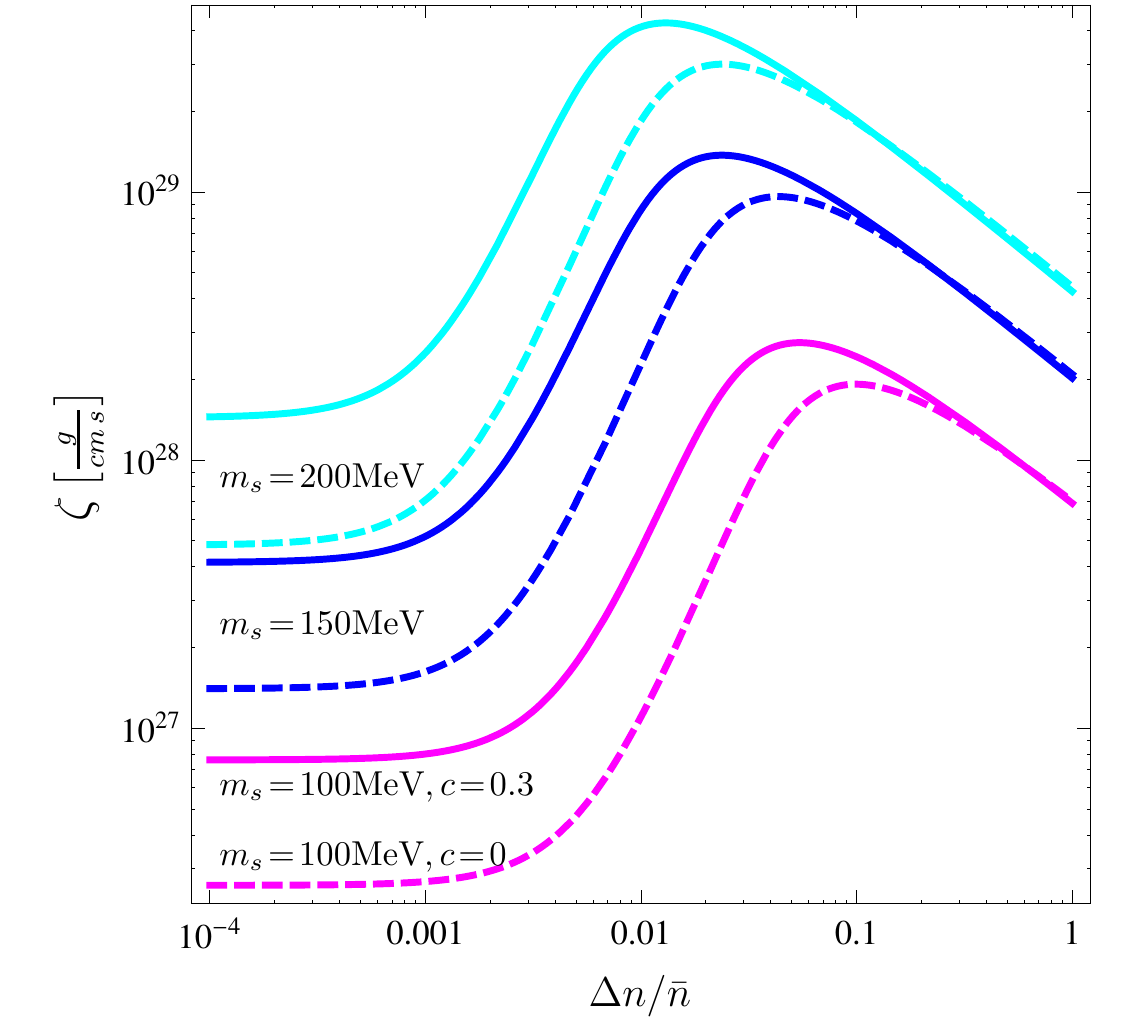}
%
%
\caption{\label{fig:quark-interaction}
The dependence of the viscosity on
parameters of the equation of state of strange quark matter using the simple
parameterization eq.~(\ref{eq:quark-eos-parametrization}). 
We show the amplitude dependence at $T=10^{8}$ K for $\omega=8.4$
kHz and $\bar n=2n_0$. Dashed curves are for $c=0$,
solid curves are for $c=0.3$. We show
$m_s=100$\,MeV (lowest two curves,
magenta online),
$m_s=150$\,MeV (middle two curves, blue online) and $m_s=200$\,MeV (highest two curves, cyan online).
}
\end{figure}

The bulk viscosity is sensitive (via the susceptibilities) to the
parameters $c$ and $m_{s}$, but not to the bag constant.  We show in
fig.~\ref{fig:quark-interaction} the effect on the bulk viscosity of
varying $c$ and $m_{s}$ within their expected range of values, at
twice nuclear saturation density and a temperature $T=10^{8}$~K.  We
calculate the bulk viscosity for an angular frequency of the oscillation of
$\omega=8.4$~kHz (corresponding to the r-mode of a pulsar with a
period of $1$~ms). We find that the uncertainty amounts to more than an order of
magnitude. In contrast to the equilibrium composition of strange stars
which proved to be strongly dependent on the parameter $c$
\cite{Alford:2004pf}, in the present case the effective strange quark
mass has a larger impact.

\begin{figure}
\includegraphics[scale=0.8]{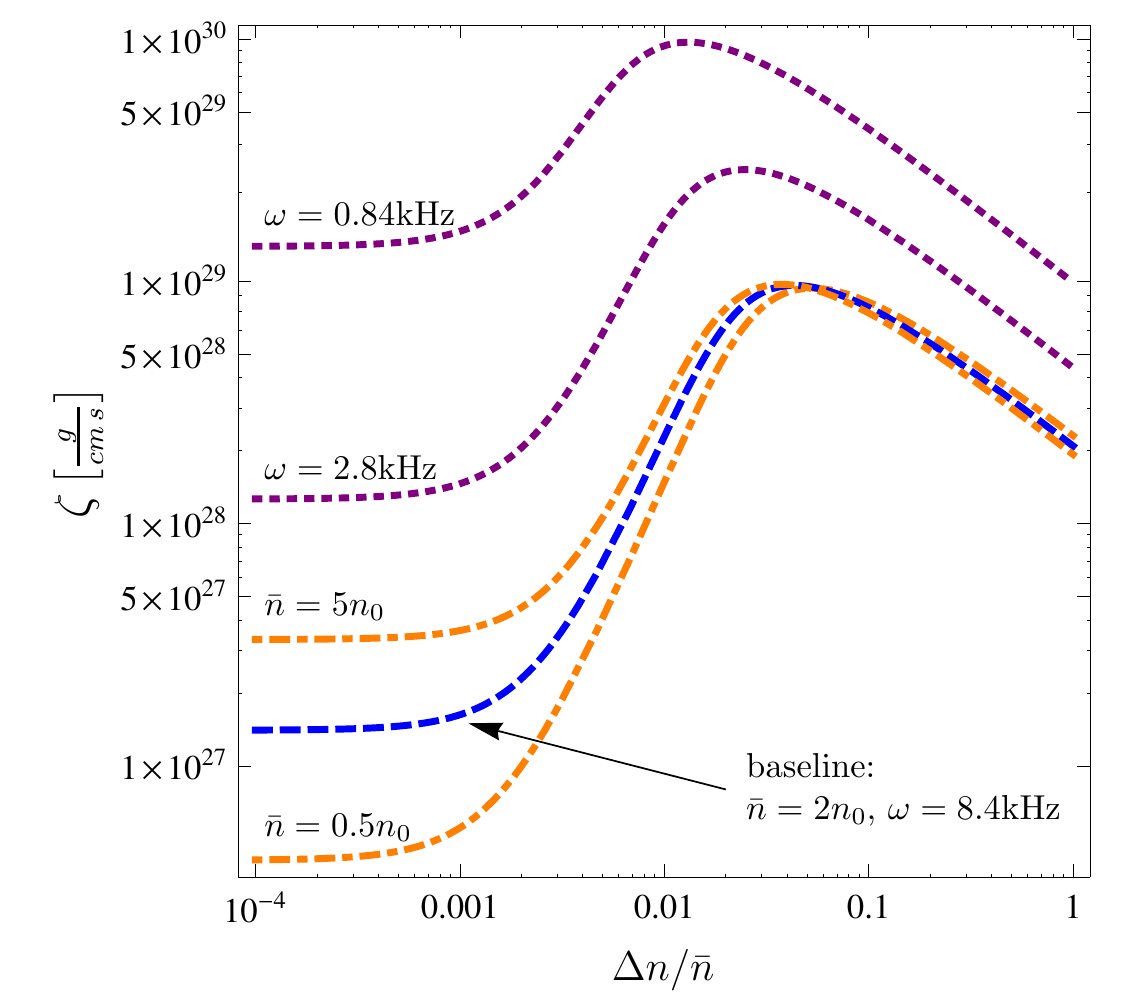}
%
%
\caption{\label{fig:density-frequency}
The dependence of the viscosity of quark matter on the 
density and oscillation frequency, using the phenomenological
equation of state eq.~(\ref{eq:quark-eos-parametrization})
with  $m_s=150$\,MeV and $c=0$, at $T=10^{8}$\,K.
The dashed (blue online) ``baseline'' curve is for $\bar n=2n_0$ and a high 
angular frequency $\omega=8.4$ kHz corresponding to a millisecond pulsar.
The dot-dashed (orange online) curves show the variation from the baseline
with density: a low value $\bar n=0.5n_0$ and a high value of $\bar n=5n_0$. The dotted 
(purple online) curves give the variation from the baseline
with angular frequency: a lower value $\omega=0.84$\,kHz and 
an intermediate value $\omega=2.8$\,kHz.
}
\end{figure}

Finally we show in
fig.~\ref{fig:density-frequency} the dependence of the viscosity of
a quark gas on the density of the matter and the
frequency of the oscillation. The
density dependence is most pronounced in the sub-thermal regime and
becomes basically irrelevant in the supra-thermal regime, in accordance with the density-independence of the maximum of the viscosity eq.~\eqn{eq:strange-maximum}.
Further, we see that the viscosity increases strongly with frequency, according to the $1/\omega$-dependence of the maximum eq.~\eqn{eq:strange-maximum} which arises as a prefactor in eq.~\eqn{eq:viscosity-parametrization}.
Therefore, the results for a millisecond-pulsar given here in all other figures
present a lower limit for the viscosity, whereas the damping of slower
rotating stars is much faster.

\section{Hadronic matter}
\label{sec:hadronic-matter}

\subsection{General features}
\label{sec:hadronic-general}

The bulk viscosity has been calculated for various phases of nuclear matter
(unpaired, superfluid, kaon-condensed etc) with flavor equilibration
via either direct or modified Urca processes 
\cite{Sawyer:1989dp,Haensel:2000vz,Haensel:2001mw-ADS,
Chatterjee:2007qs,Gusakov:2007px,Chatterjee:2007ka}. 
The leptonic contribution has recently been calculated \cite{Alford:2010jf},
and hyperonic matter has also been studied
\cite{Lindblom:2001hd,Jones:2001ya,Haensel:2001em-ADS,Chatterjee:2007iw-ADS}.
We concentrate on the simplest case of non-superfluid hadronic
$npe$ matter. We note however, that the generic properties of our
results also apply to more complicated forms of matter like hyperonic
and/or superfluid nuclear matter. In the case of hadronic matter
we assume that weak equilibration occurs via the Urca channel
\begin{equation}
p+e^{-}\to n+\nu_{e}\quad,\quad n\to p+e^{-}+\bar{\nu}_{e}\end{equation}
There are two qualitatively different cases depending on whether the
direct process is possible or only the modified version where a bystander
nucleon is necessary to satisfy energy-momentum conservation. The
latter represents a particular strong interaction vertex correction
to the above process. However, from the point of view of the weak
interaction these different processes belong to the same channel.
Taking into account baryon number and charge conservation $\delta n_{p}-\delta n_{e}=0$,
the driving baryon number density oscillation yields here the oscillating
chemical potential difference
\begin{equation}
\mu_{I}\equiv\mu_{n}-\mu_{p}-\mu_{e} \ .
\end{equation}
where the notation reflects that the equilibrating quantity in this case is isospin.

Taking into account the effect of supra-thermal oscillation amplitudes
requires the non-linear $\mu_I$-corrections to the corresponding
rates. These have been given for hadronic matter in \cite{Reisenegger:1994be,Sawyer:1989dp,Friman:1978zq}
in the standard case that only modified Urca processes are allowed
\begin{align}
 & \Gamma^{(\leftrightarrow)}_{hm} \mu_I=- 3.5\cdot10^{13}\frac{\rm ergs}{{\rm cm}^{3}{\rm s}}
  \left(\frac{x\, n}{n_{0}}\right)^{\frac{1}{3}}\frac{T_{8}^{8}}{11513}\label{eq:modified-rate}\\
 & \cdot\left(14680 \frac{\mu_I^2}{\pi^2 T^2} +7560 \frac{\mu_I^4}{\pi^4 T^4}+840\frac{\mu_I^6}{\pi^6 T^6} +24\frac{\mu_I^8}{\pi^8 T^8} \right)\nonumber \end{align}
and in the enhanced case when direct Urca processes 
dominate \cite{Reisenegger:1994be,Haensel:1992zz}
\begin{align}
 & \Gamma^{(\leftrightarrow)}_{hd} \mu_I=- 4.3\cdot10^{21}\frac{\rm ergs}{{\rm cm}^{3}{\rm s}}\left(\frac{x\, n}{n_{0}}\right)^{\frac{1}{3}}\frac{T_{8}^{6}}{457}
\label{eq:direct-rate}\\
 & \qquad\quad\cdot\left(714\frac{\mu_I^2}{\pi^2 T^2}+420\frac{\mu_I^{4}}{\pi^4 T^4}+42\frac{\mu_I^{6}}{\pi^{6}T^{6}}\right)\nonumber \end{align}
where $T_{8}$ is the temperature in units of $10^{8}$ K. Here we use the expressions given in \cite{Reisenegger:1994be}, but we note that the hadronic rates depend on model assumptions for the behavior of the strong interaction at high density (see \cite{Friman:1978zq,Haensel:1992zz}) and thereby involve uncertainties.
These expressions
yield the parameter values given in table~\ref{tab:weak-parameters}. There
are major differences between these hadronic rates and the corresponding
one for strange quark matter.
In quark matter, non-leptonic
processes are naturally allowed and only particles that have a Fermi surface
(quarks in this case) are involved.
In contrast in hadronic matter such processes are absent
(unless hyperons are present) and equilibration must proceed via
semi-leptonic processes involving particles with no Fermi surface
(neutrinos in this case) giving a much stronger temperature
dependence. As noted before the simple analytic approximation suitable
for strange quark matter is not applicable here. Nevertheless we will
see that many qualitative aspects of that solution obtain in the
general case. We note that although the prefactors of the non-linear
terms decrease strongly as the power of $\mu_I/T$ rises,
it is not sufficient to neglect them since
they enter the non-linear differential equation (\ref{eq:general-diff-eq})
where they dominate at sufficiently large amplitudes.

According to eq.~(\ref{eq:susceptibilities}), the susceptibilities
for hadronic matter are
\begin{align}
C_{h} & =\bar{n}\left(
  \left.\frac{\partial\mu_{n}}{\partial n}\right|_{x_n}
 -\left.\frac{\partial\mu_{p}}{\partial n}\right|_{x_n}
 -\left.\frac{\partial\mu_{e}}{\partial n}\right|_{x_n} \right)\,,\\
B_{h} & =
 \frac{1}{\bar{n}}\left(
   \left.\frac{\partial\mu_{n}}{\partial x_{n}}\right|_{n}
  -\left.\frac{\partial\mu_{p}}{\partial x_{n}}\right|_{n}
  -\left.\frac{\partial\mu_{e}}{\partial x_{n}}\right|_{n} \right)\,.
\end{align}
Computing these quantities requires the equation of state
of dense neutron matter.
We will perform calculations using two model equations of state
of nuclear matter. The first
one is the ``hadron gas'', consisting of an electrically
neutral beta-equilibrated mixture of free neutrons, protons, and electrons.
The second one is ``APR hadron matter'', using the well-known
model by Akmal, Pandharipande
and Ravenhall \cite{Akmal:1998cf} which relies on a 
potential model that reproduces scattering data at nuclear densities.
As a low density extension of the APR data we use \cite{Baym:1971pw,Negele:1971vb}. 
In order to make it easy to apply our general results to other equations
of state, we implement the APR equation of state
using the simple parameterization employed in \cite{Lattimer:1991ib}
to approximate the dependence of the energy per particle on the proton
fraction $x$ by a quadratic form

\begin{equation}
E(n,x)=E_{s}(n)+S(n)(1-2x)^{2}
\label{eq:quadratic}
\end{equation}
where $E_{s}$ and $S$ are the corresponding energy for symmetric
matter and the symmetry energy. We perform a global quartic fit to
the APR prediction for symmetric and pure neutron matter $E_{n}$ which then yields the symmetry energy as
\begin{equation}
S(n)=E_{n}(n)-E_{s}(n)\end{equation}
and the complete pressure including the electron contribution reads
\begin{equation}
p(n,x,\mu_{e})=n^{2}\left(\frac{d E_{s}(n)}{d n}\!+\!\frac{d S(n)}{d n}(1\!-\!2x)^{2}\right)\!+\!\frac{\mu_{e}^{4}}{12\pi^{2}}\end{equation}
In the absence of oscillations the $\beta$-equilibrium condition
$\mu_I=0$ yields the electron chemical potential as

\begin{equation}
\mu_{e}=4\left(1-2x\right)S(n)\end{equation}
and the requirement of charge neutrality $n_{p}=n_{e}$ allows us to
determine the proton fraction $x(n)$ so that the pressure
becomes a function of the baryon density alone.
With these explicit expressions the susceptibilities in table \ref{tab:strong-parameters} can be computed and the general results in section \ref{sec:general-viscosity} can be employed. 
In the following subsections we will discuss the numerical results for
the bulk viscosity of nuclear matter, comparing it with those for
one particular model of quark matter, the one given by
eq.~(\ref{eq:quark-eos-parametrization}) with 
$m_s=150$ MeV and $c=0.3$.

\subsection{Sub-thermal case}

\begin{figure}
\includegraphics[scale=.8]{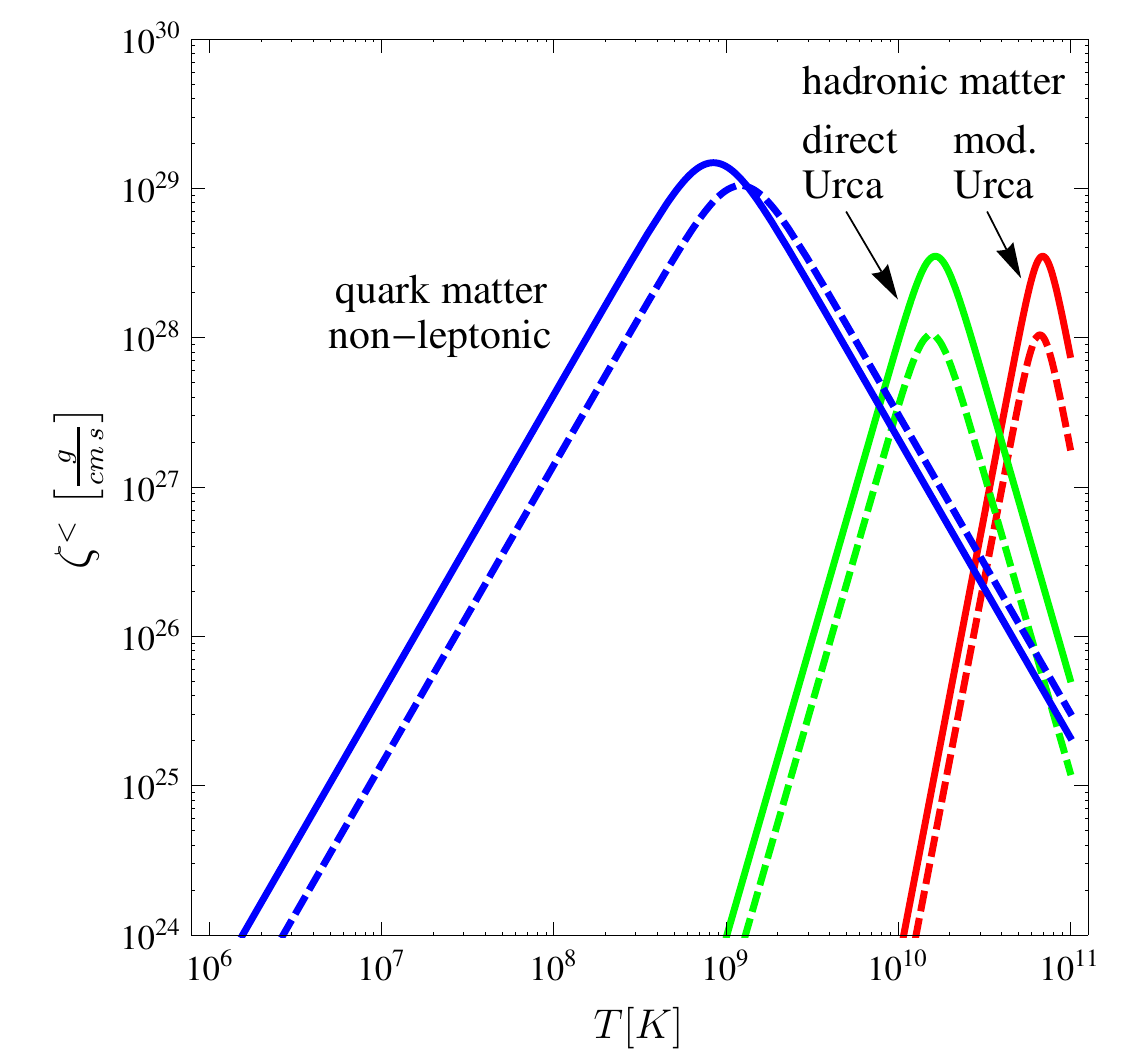}
\caption{\label{fig:subthermal-viscosity}
The sub-thermal approximation to
the viscosity at vanishing amplitude as a function of temperature
for $\omega=8.4$ kHz and $n=2n_0$. The right (red) curves
represent the standard case of hadronic matter with modified Urca
processes, the middle (green) curves hadronic matter when direct Urca
process are allowed, and the left (blue) curves strange quark matter
with non-leptonic processes. The dashed curves are for the free hadron
and free quark models; the solid curves are for APR hadron matter,
and interacting quarks eq.~\eqn{eq:quark-eos-parametrization}
with $m_s=150$\,MeV and $c=0.3$.
With APR nuclear matter the bulk viscosity
is $\sim 3$ times larger than for the free hadron gas used e.g.~in \cite{Sawyer:1989dp,Cutler:1990,Jaikumar:2008kh}.}
\end{figure}

When $\mu_\Delta \ll T$
we obtain from the analytic expression eq.~(\ref{eq:sub-viscosity})
the results shown in fig.~\ref{fig:subthermal-viscosity} where the 
bulk viscosity of strange quark matter discussed in the previous section is also
included for comparison. Here and in the following plots we
study matter at twice nuclear saturation density, $\bar{n}=2n_{0}$, 
and a compression cycle with a high angular frequency $\omega=8.4$~kHz 
corresponding to an r-mode in a pulsar with a period
of $1$~ms. We see in fig.~\ref{fig:subthermal-viscosity} that
the maximum bulk viscosity of hadronic matter as a function
of temperature (or equivalently as a function of angular frequency) is 
roughly an order of magnitude smaller than the 
maximum value for strange quark matter. This is unrelated to 
the beta-equilibration rate: the maximum viscosity depends according to eq.~\eqn{eq:sub-maximum} on the
relevant susceptibilities of the matter in question.

Other features of the plot do depend on the equilibration rate.
As we expect from \eqn{eq:sub-viscosity}, quark matter achieves its
maximum viscosity at the lowest temperature, and has less suppression
at low temperatures. This is because the nonleptonic equilibration
only involves two particles in the initial and final state, 
each of which has a large Fermi momentum $\sim\mu_q$ and hence large 
phase space factors. This leads to a low $\kappa=2$ and a large value
of $\tilde\Gamma$ (table~\ref{tab:weak-parameters}).
Thus the suppression at low temperature is only $T^2$, and, according to eq.~\eqn{eq:sub-maximum}, $T_{\rm max}$
is relatively low.
The next fastest is the direct Urca process in nuclear matter, which
involves more particles (including neutrinos which have no Fermi surface
and thus very little phase space)
and therefore has a higher $\kappa$ and lower $\tilde\Gamma$, giving it
stronger $T^4$ suppression at low temperatures, and a higher $T_{\rm max}$.
The slowest is the modified Urca process in nuclear matter, which
involves additional spectator nucleons, raising $\kappa$ to 6 and
further lowering  $\tilde\Gamma$, raising $T_{\rm max}$, and
increasing the low-$T$ suppression to $T^6$.

Note that the right-most solid and dashed curves in 
fig.~\ref{fig:subthermal-viscosity}, for hadronic matter with
modified Urca equilibration, correspond roughly to the
leftmost of the three solid (red online) curves in fig.~\ref{fig:gen-sol}(a)
that run along the surface from front to back.

We draw two important conclusions from fig.~\ref{fig:subthermal-viscosity}.
First, we have retained the full resonant structure of the viscosity
compared to previous analyses \cite{Sawyer:1989dp,Haensel:1992zz}
where a low temperature approximation $\tilde\Gamma B T^\kappa\ll\omega$ was used. 
This allows us to see that the viscosity decreases again at large temperatures and the maximum
(\ref{eq:sub-maximum}) occurs at millisecond-scale
frequencies at potentially physically relevant
temperatures of the order $10^{10}$~K for direct Urca and
$10^{11}$~K for modified Urca. This means that the resonant structure
may be important in some astrophysical applications and from
eq.~(\ref{eq:sub-maximum}) it is clear that it becomes increasingly important at
lower frequencies. 
Second, we see in fig.~\ref{fig:subthermal-viscosity} that 
for nuclear matter there is a considerable difference between
the solid curves which are based on an interacting equation of state
\cite{Akmal:1998cf,Baym:1971pw,Negele:1971vb}
and the dashed curves which are for a free gas\footnote{Note that strictly speaking there are no modified Urca processes in an ideal hadron gas. Yet, for comparison with previous studies we use here the interacting matter expression for the rate but the ideal gas expressions for the strong susceptibilities.} of nucleons and electrons.
These models have different susceptibilities $B$ and $C$, and the main effect
of this is a vertical shift of the whole curve. The shift in $T_{\rm max}$
is smaller because of the square root in eq.~(\ref{eq:sub-maximum}).
Hadronic matter with interactions has
been considered (with a more simplified equation
of state) in \cite{Haensel:1992zz,Haensel:2000vz} but many analyses
\cite{Lindblom:1998wf,Owen:1998xg,Jaikumar:2008kh}
rely on the simple analytic result%
\footnote{Note that the numerical prefactor given in \cite{Sawyer:1989dp}
is too large by two orders of magnitude, see also \cite{Cutler:1990}.%
} given by Sawyer \cite{Sawyer:1989dp} which is based on the free
gas expression. We see that these differ by roughly a factor of three for the given density of $\bar n=2n_0$, but according to fig. \ref{fig:max-viscosity} this difference can increase strongly both at lower and higher density.

\subsection{The supra-thermal regime}

Beyond the sub-thermal limit, a numeric evaluation of 
eqs.~(\ref{eq:general-diff-eq})
and (\ref{eq:general-solution}) is required. As discussed 
in sect.~\ref{sec:general-viscosity},
the temperature and amplitude dependence of the bulk viscosity 
for a given form of matter can be expressed in terms of the
function ${\cal I}(d,f)$ which was plotted in a form that is independent of the equation of state 
of hadronic matter with modified Urca processes
in fig.~\ref{fig:gen-sol}. 

\begin{figure*}
\begin{minipage}[t]{0.5\textwidth}%
\begin{center}(a)~\underline{$T=10^6$~K}\end{center}
\includegraphics[scale=.8]{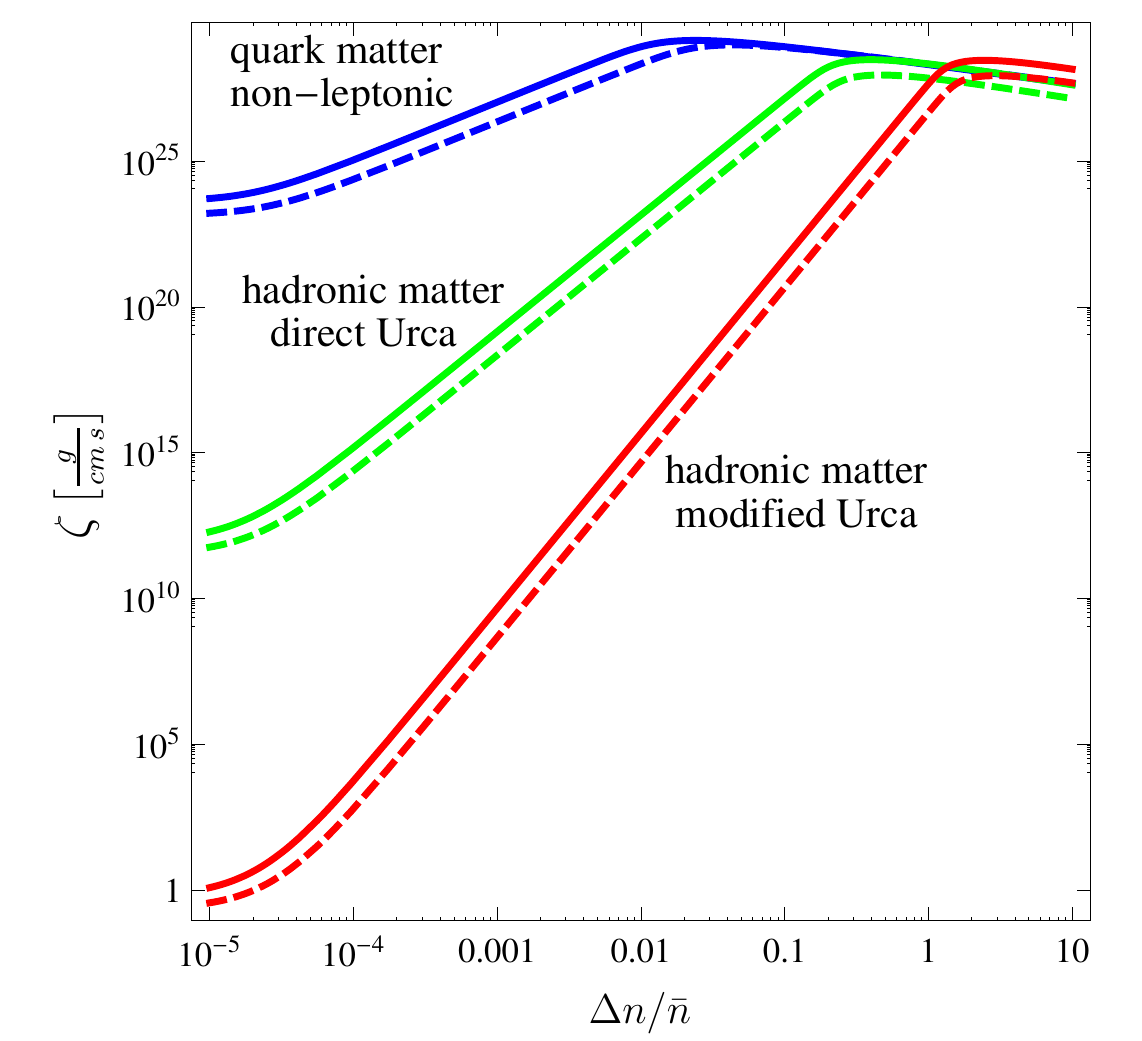}%
\end{minipage}%
\begin{minipage}[t]{0.5\textwidth}%
\begin{center}(b)~\underline{$T=10^{9}$~K}\end{center}
\includegraphics[scale=.8]{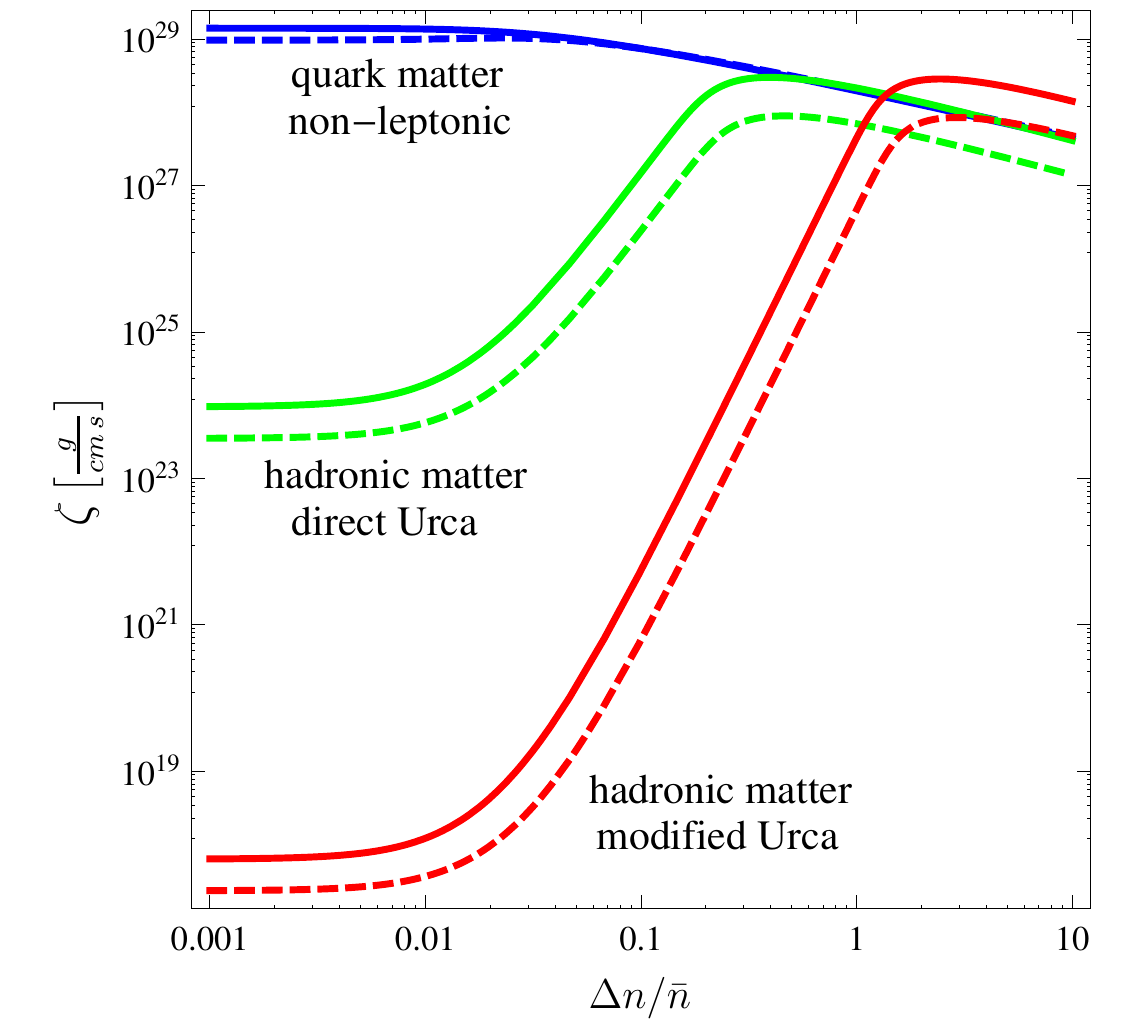}%
\end{minipage}

\caption{\label{fig:all-viscosity} 
Comparison of the bulk viscosity of the different
forms of matter studied in this work as a function of the
density oscillation amplitude $\Delta n/\bar{n}$. The frequency is
$\omega=8.4$~kHz, corresponding to an r-mode in a millisecond pulsar and $\bar n = 2n_0$.
Left panel:
low temperature $T=10^{6}$~K; right panel: high temperature $T=10^{9}$~K.
The dashed curves are for the free hadron
and free quark models; the solid curves are for APR hadron matter,
and interacting quarks eq.~\eqn{eq:quark-eos-parametrization}
with $m_s=150$\,MeV and $c=0.3$.
The bottom (red online) curves
represent the standard case of hadronic matter with modified Urca
processes, the middle (green online) curves are for hadronic matter when direct Urca
process are allowed and the top (blue online) curves are for strange quark matter
with non-leptonic processes. 
Our calculations are valid only for $\Delta n/\bar{n}\ll 1$, but we show their
extrapolation to higher amplitudes in order to compare with the qualitative general structure of the solution in fig.~\ref{fig:gen-sol}. Note that this plot uses a high oscillation frequency and that the viscosity is even larger at smaller values.
}
\end{figure*}

Using this result, we show in fig.~\ref{fig:all-viscosity} 
plots of the amplitude dependence of the bulk viscosity
at two temperatures (left panel: $T=10^6$~K; right panel: $T=10^{9}$~K)
for the various forms of hadronic and quark matter considered in this paper.
Here solid lines again show
the results for interacting matter whereas the dashed lines show the
free hadron/quark gas results. 

Note that the right-most curves in 
\ref{fig:all-viscosity}, for hadronic matter with
modified Urca equilibration, corresponds roughly to the
foremost of the three dashed (blue online) curves in fig.~\ref{fig:gen-sol}(a)
that run along the surface from left to right.

At the lower temperature the viscosity reaches
the supra-thermal regime already for small amplitudes, whereas 
at the higher temperature the
sub-thermal regime extends to large amplitudes, giving a
flat amplitude-independent plateau at low amplitudes.
The stronger non-linear feedback
in the hadronic cases leads to a significantly steeper rise that correlates
with the largest power in eq.~(\ref{eq:gamma-parametrization}). Interestingly,
despite these differences the maximum value reached by varying the
amplitude is still roughly the same as the maximum value in the sub-thermal
limit eq.~(\ref{eq:sub-maximum}), as has been analytically found
in the case of strange quark matter. This is important since it means
that oscillations are approximately equally 
damped at all temperatures once
the amplitude becomes sufficiently large. The maximum arises for amplitudes
of the order $0.01$, $0.1$ and $1$ for strange quark matter and
hadronic matter with direct and modified Urca, respectively.
The supra-thermal enhancement of the bulk viscosity is
so strong, particularly for hadronic matter,
that it could well provide the main saturation
mechanism for unstable r-modes, stopping their growth at amplitudes
that are below the threshold for other competing saturation mechanisms
(e.g. non-linear hydrodynamics) but large enough to allow 
spin-down of a neutron star
via gravitational radiation on astrophysical time scales.

%

\section{Conclusions}

We have studied the bulk viscosity of dense matter including its non-linear
behavior at large amplitudes. In particular we give a general solution
for the bulk viscosity of degenerate matter that is valid for 
arbitrary equations of state
and retains its full parameter dependence in sec.~\ref{sec:general-viscosity}. 
This allows one to include
these supra-thermal effects in a systematic r-mode analysis. In the
supra-thermal regime we give a general analytic result for strange quark matter with non-leptonic processes in sec.~\ref{sub:Strange-quark-matter}.
We found that the free hadron gas model of nuclear matter,
used for example in \cite{Lindblom:1998wf,Owen:1998xg,Jaikumar:2008kh}
to compute the susceptibilities that enter the viscosity,
is not accurate even in the sub-thermal regime, and 
may significantly underestimate the viscosity. 
Moreover, we find that the
standard low temperature (high frequency) approximation
is not applicable for temperatures around $10^{10}$~K
and the full resonant form of the bulk viscosity is required. 
We confirm previous results for the amplitude-dependence
of the bulk viscosity of strange quark matter \cite{Madsen:1992sx}
and find that these supra-thermal effects are parametrically even more important
in nuclear matter, because of higher-order non-linearities
in the amplitude-dependence of the Urca rate.

The most obvious application of our results is to the damping
of unstable r-mode oscillations in neutron stars. As the amplitude
of the mode enters the supra-thermal regime
the viscosity will increase steeply above the
sub-thermal result and can exceed it by many orders of magnitude,
but eventually it reaches an upper bound that
is completely independent of the particular weak damping process
and depends only on susceptibilities of the dense matter in question.
The viscosity then decreases at even larger amplitudes. 
We conclude that if r-mode growth is not stopped by the supra-thermal
bulk viscosity before this maximum is reached then
other non-linear dynamic
effects \cite{Lindblom:2000az,Gressman:2002zy,Lin:2004wx,Bondarescu:2007jw}
will be required to stop it. 
We have already performed initial exploratory 
calculations of r-mode damping times, and these
suggest that over a significant region of parameter space
supra-thermal bulk viscosity is sufficient to saturate r-mode growth
at a finite amplitude parameter $\alpha_{max}<1$,
as was previously assumed \cite{Owen:1998xg}.
This topic will be
discussed in more detail in a forthcoming publication. 
There are several other directions in which our research could be
developed. Other equations of state for quark matter could be
studied,
for example the perturbative equation of state \cite{Kurkela:2009gj},
and also other phases with different equilibration mechanisms.
The same is true for the various equations of state
and phases of hadronic matter. Our analysis was for the case of a
single equilibration channel, and it would be interesting to
extend it to multiple channels, which may be relevant to both hadronic 
and quark matter (see appendix A of \cite{Alford:2006gy} and 
Ref.~\cite{Sa'd:2007ud}). 
In quark matter the non-Fermi liquid enhancement of the Urca rate 
\cite{Schafer:2004jp}
should further increase resonant effects.
The application of our results to r-modes in neutron stars also
raises interesting questions concerning the correct treatment of
the crust \cite{Steiner:2007rr}, and possible
modification of the radial profile of the r-mode due to strong
radial dependence of the bulk viscosity in layered stars
such as hybrid stars.

Finally we note that at low temperatures the suprathermal enhancement of 
bulk viscosity becomes large, and the amplitude threshold for entering
the suprathermal regime becomes low.
This may not be relevant to the damping of
r-modes because they are also damped by shear viscosity which becomes large at
low temperature. But for other modes of compact stars, such as
monopole pulsations, shear viscosity will not play such a significant
role, and suprathermal bulk viscosity might be the dominant source
of damping if external perturbations make the amplitude large enough.
This might be relevant to old, cold, accreting stars in binary systems.

\begin{acknowledgments}
We thank Prashanth Jaikumar, Andreas Schmitt, Igor Shovkovy and Andrew Steiner
for helpful discussions. This research was supported in part by the
Offices of Nuclear Physics and High Energy Physics of the
U.S. Department of Energy under contracts
\#DE-FG02-91ER40628,  
\#DE-FG02-05ER41375. 
\end{acknowledgments}

\renewcommand{\href}[2]{#2}

\newcommand{\apjl}{Astrophys. J. Lett.\ }
\newcommand{\mnras}{Mon. Not. R. Astron. Soc.\ }
\newcommand{\aap}{Astron. Astrophys.\ }

\bibliographystyle{JHEP_MGA}
\bibliography{supra}

\end{document}